\documentclass[12pt,reqno]{amsart}

\usepackage{graphicx}
\usepackage{amsthm,amsmath,amsfonts,amssymb,amsxtra,appendix,bookmark,dsfont,bm,color, braket}
\usepackage{cite}
\usepackage{mathtools}
\usepackage[truedimen,top=3truecm,bottom=3truecm,left=2.5truecm,right=2.5truecm]{geometry}

\newtheorem{theorem}{Theorem}[section]
\newtheorem{lemma}[theorem]{Lemma}

\theoremstyle{definition}

\newtheorem{remark}[theorem]{Remark}

\newtheorem{proposition}[theorem]{Proposition}
\newtheorem{corollary}[theorem]{Corollary}
\newcommand{\C}{\mathbb{C}}
\renewcommand{\epsilon}{\varepsilon}

\renewcommand{\phi}{\varphi}
\newcommand{\R}{\mathbb{R}}
\newcommand{\s}{\mathbb{S}}
\newcommand{\til}{\widetilde}
\newcommand{\<}{\langle}
\renewcommand{\>}{\rangle}
\numberwithin{equation}{section}

\begin{document}

	\title[Exponential Lower Bounds for the Averaged Density]{Exponential Lower Bounds for the Spherically Averaged Density of Atoms}
	\author[Y. Goto]{Yukimi Goto}
	\address{Graduate School of Mathematical Sciences, University of Tokyo, Komaba, Meguro-ku, Tokyo 153-8914,
		Japan}
	\email{yukimi@ms.u-tokyo.ac.jp}

	\begin{abstract}
		We study the asymptotic behavior of the spherically averaged density $\til \rho$ for fermionic bound states of the $N$-electron atomic Hamiltonian $H_{N,Z}$ with nuclear charge $Z$.
		For the density associated with an eigenfunction satisfying $H_{N,Z}\psi=E\psi$, we show that if the energy $E$ is below the bottom of the essential spectrum $\Sigma:=\inf\mathop{\rm ess.spec} (H_{N,Z})$, then
		\[
		\lim_{r\to\infty}\frac{\ln\til \rho(r)}{r} =\lim_{r\to\infty}\frac{d}{dr}{\ln\til \rho(r)} =-2\alpha_0,
		\]
		with $\alpha_0=\sup\{\alpha \ge 0\colon e^{2\alpha r}\til \rho \in L^1(\R_+,r^2dr)\}$.
		The same is true at threshold in certain cases.
		Moreover, we prove $\alpha_0^2=\Sigma-E$ for the ground states of $H_{3,Z}$ with sufficiently large $Z$.
	\end{abstract}
	\maketitle

	\section{Introduction}
	Studying the asymptotic behavior of solutions to the $N$-particle Schr\"odinger equation is an important and classical issue, and there is a large literature devoted to its analysis. See~\cite{Agmon,AHO,AHOM,CT,DHSV,HH} and references therein.
	In contrast to the problem of establishing upper bounds, sharp lower bounds are less well understood.
	For the positive ground state $\psi$ below the essential spectrum, Carmona and Simon~\cite{CS} proved
	\begin{equation}
	\label{eq.CS}
	\lim_{|x|\to\infty}\frac{-\ln \psi(x)}{\varrho(x)}=1
	\end{equation}
	with $\varrho$ the Agmon distance~\cite{Agmon}.
	For some particular systems, more precise asymptotic results were obtained, e.g., in~\cite{AHOM,HO79,LS,HJL}.
	Unfortunately, a pointwise lower bound like~\eqref{eq.CS} is hopeless for general eigenstates, because their nodal sets can be unbounded.

	On the other hand, Bardos and Merigot~\cite{BM} proved similar lower bounds for the spherical average of the squared modulus of an eigenstate, $|\psi|^2$, in a one-body problem.
	This quantity is natural because, if the energy is a discrete eigenvalue, the spherical average is strictly positive for all large $|x|$.
	 Their method was generalized to $N$-body Hamiltonians by Froese and Herbst~\cite{FH}.
	For one-body Schr\"odinger operators, Herbst~\cite{HerbstCones} also established sharp exponential decay rates in cones.

	 	In this paper we investigate the asymptotic behavior of the spherically averaged one-electron density $\til \rho$, instead of $\psi$.
	 One important point is that the spherical average considered in~\cite{FH} is taken in the full configuration space, while $\til \rho$ is obtained by integrating out $N-1$ electrons and averaging the remaining variable over $\s^2$.
	 Fournais et al.~\cite{FHS2} proved an exponential lower bound for $\til \rho$ by extending the result of Froese and Herbst~\cite{FH}, but the upper and lower bounds have different rates.
	 Our goal here is to improve their result.
	 In particular, we establish upper and lower bounds with the same exponential rate.

	 We proceed to formulate our results precisely.
	 The $N$-particle atomic Hamiltonian is given by
	\[
	H_{N,Z}=-\sum_{i=1}^N\Delta_i+V_N(\underline{X}),\quad
	V_N(\underline{X}):=-\sum_{i=1}^N\frac{Z}{|x_i|}+\sum_{1\le i<j\le N}|x_i-x_j|^{-1},
	\]
	acting as a self-adjoint operator on $\bigwedge^N L^2(\R^3; \C^2)$, which is the subspace consisting of all antisymmetric functions; $\psi \in \bigwedge^N L^2(\R^3; \C^2)$ obeys
	\[
	\psi(\dots, z_i,\dots, z_j,\dots)=-\psi(\dots, z_j,\dots, z_i,\dots),\quad z_i:=(x_i,\sigma_i) \in \R^3\times \{\uparrow, \downarrow\}.
	\]
	Here $Z>0$ is the nuclear charge, $\Delta_i$ denotes the three-dimensional Laplacian with respect to $x_i \in \R^3$, and $\underline{X}=(x_1,\dots,x_N)$.
	We write $r_i:=|x_i|$ below.
	In addition, $\rho$ denotes the one-electron (spin-summed) density of $\psi$ defined by
	$$
	\rho(x_N):=N\sum_{\underline{\sigma}}\int_{\R^{3N-3}}|\psi(z_1,\dots,z_N)|^2\,d\hat{x}_N,\quad d\hat{x}_N:=dx_1\cdots dx_{N-1},
	$$
	where $\hat x_i:=(x_1,\dots,x_{i-1},x_{i+1},\dots,x_N)$ denotes the vector in $\R^{3N-3}$ obtained by removing $x_i$ from $\underline{X}$, and
	\[
	\sum_{\underline{\sigma}}:=\sum_{\sigma_1=\uparrow,\downarrow} \cdots \sum_{\sigma_N=\uparrow,\downarrow}.
	\]
	For $\underline{z}=(z_1,\dots,z_N)$ we use the notation
	\[
	\int f(\underline{z}) \,d\underline{z}:=
	\sum_{\underline{\sigma}}\int_{\R^{3N}} f(\underline{z}) \,d\underline{X}.
	\]
	We also write $\hat z_i:=(z_1,\dots,z_{i-1},z_{i+1},\dots,z_N)$ for the vector obtained by omitting $z_i$ from $\underline{z}$, and use the convention
	\[
	\int f(\hat z_N)\,d\hat z_N:=
	\sum_{\sigma_1=\uparrow,\downarrow}\cdots\sum_{\sigma_{N-1}=\uparrow,\downarrow}
	\int_{\R^{3N-3}}f(\hat z_N)\,d\hat{x}_N.
	\]
	In addition, we consider the {\it spherically averaged density}
	\[
	\til{\rho}(r):=\int_{\s^2} \rho(x)\,d\omega,\qquad r=|x|,\quad\omega=\frac{x}{r} \in \s^2.
	\]
	By~\cite[Theorem~1.1]{FHSanalytic}, $\til{\rho}$ is real analytic on $(0,\infty)$.
	In~\cite{FHS1}, it was shown that $\til\rho(r)>0$ for all $r\ge0$ if  $\psi$ is a ground state decaying exponentially.
	We use the normalization
	\[
	\|\psi\|^2_2:=	\int |\psi(\underline{z})|^2 \,d\underline{z}=\sum_{\underline{\sigma}} \int_{\R^{3N}}|\psi(\underline{z})|^2\,d\underline{X}=1.
	\]

	Let $|x|_\infty = \max_{1\le i \le N}|x_i|$ denote the distance of an outermost electron from the nucleus. For an eigenstate satisfying $H_{N,Z}\psi =E \psi$, we define
	\[
	\alpha_0:=\sup\{\alpha\ge 0\colon e^{\alpha |x|_\infty}\psi \in L^2(\R^{3N};\C^{2^N})\}
	=\sup\left\{\alpha \ge 0\colon  e^{2\alpha r} \til{\rho}(r) \in L^1(\R_+,r^2dr)\right\}.
	\]
	Here we have used antisymmetry and
	\[
	 \|e^{\alpha |x|_\infty}\psi \|_2^2
	 \le \sum_{i=1}^N\|e^{\alpha |x_i|}\psi \|_2^2
	 =\int_0^\infty e^{2\alpha r}\til \rho(r)r^2\,dr
	 \le N \|e^{\alpha |x|_\infty}\psi \|_2^2.
	\]
	We also introduce
	\[
	\alpha_{\underline{X}}:=\sup\{\alpha\ge 0\colon e^{\alpha |\underline{X}|}\psi \in L^2(\R^{3N};\C^{2^N})\},\quad  |\underline{X}|=\left(\sum_{i=1}^N|x_i|^2\right)^{1/2}.
	\]
	It is clear that $\alpha_{\underline{X}} \le \alpha_0 \le \sqrt{N} \alpha_{\underline{X}}$ since
		\[\frac{	|\underline{X}|}{\sqrt{N}}\le		|x|_\infty \le 	|\underline{X}|,\]
	 and, in general $N$-body systems, $\alpha_0$ and $\alpha_{\underline{X}}$ need not coincide; see~\cite[Ex.~13]{FHS2}.
	In~\cite{FHHH}, $\alpha_{\underline{X}} \le \sqrt{|E|}$ was also proved.

	From now on, $E_m$ denotes the ground state energy of $H_{m,Z}$ given by
	\[
	E_m:=\inf {\rm spec} (H_{m,Z})=\inf_{\<\psi,\psi\>=1}\left \{\<\psi, H_{m,Z}\psi\> \colon \psi \in  \bigwedge^m L^2(\R^3;\C^2)\cap H^2(\R^{3m}; \C^{2^m})  \right\},
	\]
	and if it is an eigenvalue, the corresponding eigenstate is called the ground state.
	By the HVZ theorem~\cite{CFKS}, ${\rm ess.spec} (H_{N,Z})=[E_{N-1},\infty)$.

	Our main theorem is as follows.
	\begin{theorem}
		\label{thm.main}
		Let $\psi$ be a normalized eigenstate of $H_{N,Z}$ with eigenvalue $E<E_{N-1}$. Then we have
	\begin{equation}
		\label{eq.th}
		\lim_{r\to \infty}\frac{\ln {\til{\rho}(r)}}{r} =\lim_{r\to \infty} \frac{d}{dr} \ln \til{\rho}(r)=-2\alpha_0.
		\end{equation}

		Moreover, \eqref{eq.th} is true for eigenstates with eigenvalue $E=E_{N-1}<E_{N-2}$ and $Z<N-1$.
\end{theorem}

We emphasize that the result covers fermions and does not require positivity of wave functions.

\begin{remark}
		This theorem is also valid for particles with any finite number $q$ of spin states: $\psi \in \bigwedge^N L^2(\R^3; \C^q)$.
		In addition, if $E_N=E_{N-1}$ at a critical nuclear charge $Z_c\in (N-2,N-1)$, $E_{N-1}<E_{N-2}$ follows from Zhislin's theorem~\cite{Zhislin} and there exists a threshold ground state~\cite{HOS,Gridnev,BFLS}.
		A more general existence result is given in~\cite{BFLS}.
		Theorem~\ref{thm.main} also extends to molecules with fixed nuclei, with adequate modifications.
	\end{remark}

\begin{remark}
	By Theorem~\ref{thm.main}, we have for any $\epsilon>0$
	\[
	c_\epsilon e^{(-2\alpha_0-\epsilon) r} \le \til \rho(r)\le C_\epsilon e^{(-2\alpha_0+\epsilon) r}
	\]
	for sufficiently large $r$.
	On the other hand, it is known~\cite{AHOM} that the one-electron density $\rho$ satisfies
	\[
	\rho(x)\le C|x|^{\frac{Z-N+1}{\sqrt{E_{N-1}-E}}-2}e^{-2\sqrt{E_{N-1}-E}|x|}
	\]
	for $|x|$ large enough when $E<E_{N-1}<E_{N-2}$.
	This shows $\alpha_0\ge \sqrt{E_{N-1}-E}$.
	For ground states of two-electron or bosonic atoms, a corresponding lower bound is known~\cite{AHOM, Goto2} if $E_N<E_{N-1}<E_{N-2}$.
	Subexponential decay at threshold is also known~\cite{HOS,HJL,HJLAtoms,Goto2}.
	We expect that a similar sharp lower bound including the polynomial prefactor holds for $\til \rho(r)$ when $\psi$ is a ground state even for $N\ge 3$ fermions, but have not been able to prove it.
	\end{remark}

\begin{remark}
	In~\cite[Thm.~6]{FHS2}, it was proved that for $E<E_{N-1}$
	\begin{equation}
		\label{eq.FHS}
	-2\sqrt{N}\alpha_{\underline{X}}\le
	\liminf_{r\to \infty}\left(\frac{\ln {\til{\rho}(r)}}{r} \right)
	\le
	\limsup_{r\to \infty}\left(\frac{\ln {\til{\rho}(r)}}{r} \right)\le-2
\alpha_{\underline{X}}.
	\end{equation}
	\end{remark}

	For general $N$-particle systems, the possible difference between $\alpha_0$ and $\alpha_{\underline{X}}$ and the appearance of the factor $\sqrt{N}$ were discussed in~\cite{FHS2}, but we prove equality for ground states of three-electron atoms with sufficiently large $Z$.

	\begin{theorem}
		\label{thm.asymp}
		Suppose $N=3$ and $\psi$ is a ground state.
		For all sufficiently large $Z$, we have
		\[
		\alpha_0=\alpha_{\underline{X}} =\sqrt{E_2-E_3}.
		\]
		\end{theorem}

		\begin{remark}
			In~\cite{FH82}, Froese and Herbst proved that $\alpha_{\underline{X}}^2+E \in\mathcal{F}(H_{N,Z})$ provided $\alpha_{\underline{X}}<\infty$, where the set $\mathcal{F}(H_{N,Z})$ denotes the thresholds of $H_{N,Z}$, i.e., $\lambda \in \mathcal{F}(H_{N,Z})$ if and only if either $\lambda=0$ or $\lambda$ is an eigenvalue of a proper subsystem of $H_{N,Z}$.
			Then $\alpha_{\underline{X}}^2=E_{N-1}-E$ means that $E+\alpha_{\underline{X}}^2$ coincides with the lowest threshold.
			\end{remark}

			We prove Theorem~\ref{thm.main} in Section~\ref{sect.main}.
			The main difficulty is to treat $\alpha_0$ rather than $\alpha_{\underline{X}}$.
			Froese and Herbst~\cite[Thm.~2.1]{FH} (see Thm.~\ref{thm.FH} below) obtained strong information on $\alpha_{\underline{X}}$ using a Mourre estimate, and in~\cite{FHS2}, Fournais et al. used the characterization to obtain \eqref{eq.FHS}.
			In contrast, it is difficult to apply the Mourre estimate to $\alpha_0$, although $\alpha_0$ is natural for studying the asymptotics of $\til \rho$.
			To overcome this difficulty, we use the comparison method for Schr\"odinger equations developed in~\cite{HH,HO79,AHOM}.
			In order to reduce the problem to a differential inequality, we use a virial-type identity for $\til \rho$ in Lemma~\ref{lem.virial}.
			We then prove the convexity of a suitably corrected logarithm of $\til\rho$, which yields the desired asymptotics.
			Related log-convexity was investigated in~\cite{AD}.
			From a technical point of view, the use of this kind of comparison principle to obtain log-convexity is a novel feature of our analysis.

			Section~\ref{sect.largeZ} is devoted to the proof of Theorem~\ref{thm.asymp}.
			The proof relies heavily on perturbation theory and the structure of hydrogen atoms.
			To establish the equality, we also prove a sufficient condition for $\alpha_0=\alpha_{\underline{X}}$ in Proposition~\ref{lem.equiv}.
			The result is obtained by a standard localization method, but it allows us to use the threshold characterization of~\cite{FH82}.
			Finally, Friesecke and Goddard~\cite{FG} explicitly determined the large-$Z$ asymptotics of low-lying atomic energy levels and eigenstates for $N\le 10$, and their analysis might be useful for extending our results.

	\section{Proof of Theorem~\ref{thm.main}}
	\label{sect.main}
Without loss of generality, we may assume any bound state $\psi$ is real-valued.

First, we refine the pointwise bounds for $\til \rho$ in~\cite{FHS2} in the following sense.

\begin{lemma}
	\label{lem.pbound}
	Assume $H_{N,Z}\psi = E\psi$.
	\begin{enumerate}
	\item Suppose $E<E_{N-1}$ and fix $0<\mu<\sqrt{E_{N-1}-E}$.
	Then, there exists $R_0>0$ such that for all $R\ge R_0$
	\begin{equation}
		\label{eqlem.discpbound}
	\frac{2\mu}{R^2}\int_R^\infty \til{\rho}(r)r^2\,dr
	\le \til{\rho}(R)
	\le
	\frac{1}{R^2}\int_{R-1}^\infty \til{\rho}(r)r^2\,dr.
	\end{equation}

	\item If $E=E_{N-1}<E_{N-2}$ with $N-1-Z>0$, then for any $0<\mu <\sqrt{N-1-Z}$ there exists $R_0>0$ such that

			\begin{equation}
			\label{eqlem.threpbound}
	\frac{2\mu}{R^{5/2}}\int_R^\infty \til{\rho}(r)r^2\,dr
	\le \til{\rho}(R)
	\le
	\frac{1}{R^2}\int_{R-1}^\infty \til{\rho}(r)r^2\,dr
\end{equation}
	for $R\ge R_0$.
	\end{enumerate}
	\end{lemma}

	\begin{remark}
		For comparison, the authors of~\cite{FHS2} proved
		\[
		\frac{c}{R^3}\int_R^\infty \til{\rho}(r)r^2\,dr
		\le \til{\rho}(R)
		\le
		\frac{C}{R^2}\int_{R-1}^\infty \til{\rho}(r)r^2\,dr
		\]
		for sufficiently large $R$.
		Our refinement is crucial for the lower bound.
		\end{remark}

		\begin{proof}[Proof of Lemma~\ref{lem.pbound}]
	The arguments in~\cite[Lemma~3.1]{AHOM} yield the following inequalities; see also~\cite[Lemma~3.1]{Goto2}.
	\begin{lemma}
		For $H_{N,Z}\psi = E\psi$, we consider the two cases:
		\begin{enumerate}
			\item Suppose $E<E_{N-1} $.
			For sufficiently large $|x_N|$,
			\begin{equation}
				\label{lemeq.Schroine}
				\left(
				-{\Delta_N}+E_{N-1}-E-\frac{Z}{r_N}
				\right)\sqrt{\rho(x_N)}\le0
			\end{equation}
			in the distributional sense.

			\item Suppose $E=E_{N-1}<E_{N-2}$ with $N-1>Z$.
			There is a positive constant $d$ depending on $N,Z$ such that for sufficiently large $|x_N|$
		\begin{equation}
		\label{lemeq.threSchroine}
		\left(
		-{\Delta_N}+\frac{N-1-Z}{r_N}-\frac{d}{r_N^2}
		\right)\sqrt{\rho(x_N)}\le0
		\end{equation}
		in the distributional sense.
		\end{enumerate}
	\end{lemma}
	First, we consider the case $E<E_{N-1}$.
	Taking the spherical average in~\eqref{lemeq.Schroine}, we see
	\[
	0\ge -\left(\frac{d^2}{d r^2} +\frac{2}{r}\frac{d}{d r} -\mu^2\right)\sqrt{\til{\rho}(r)},
	\]
	where we have used the Hoffmann-Ostenhof inequality~\cite[Eq.~(2.23)]{HH}
	\[
	|\nabla \sqrt{\til \rho (r)}|^2\le
	N\sum_{\sigma=\uparrow, \downarrow}\int_{\s^2}d\omega\int d\hat z_N\left|\frac{\partial \psi}{\partial r}(\hat z_N, z)\right|^2, \quad z=(x,\sigma).
	\]
	Now we introduce  $v(r):=r\sqrt{\til{\rho}(r)}$ so that $v''\ge \mu^2 v$.
	Since $v''\ge0$, $v \in L^2(\R_+,dr)$ is convex and nonincreasing for large $r$.
	In fact, if $v'(r_0)>0$ for some $r_0$, then $v(r)\ge v(R)+v'(R)(r-R)$ with $v'(R)>0$ for $r> R>r_0$ by the convexity, but this contradicts $v \in L^2$.
	From this, we see
	\[
	v(R)^2\le \int_{R-1}^R v(r)^2 \,dr \le \int_{R-1}^\infty \til{\rho}(r)r^2\,dr,
	\]
	 which proves the upper bound in~\eqref{eqlem.discpbound}.
	 This also leads to $v\to 0$.
	Then the maximum principle yields
	\[
	\sup_{r\ge R}(v(r)-v(R)e^{-\mu(r -R)})\le 0,
	\]
	which shows
	\begin{equation}
		\label{eq.vbound}
	v(r)\le v(R)e^{-\mu(r -R)}
	\end{equation}
	for $r\ge R$.
	Then we obtain the lower bound
	\begin{equation}
		\label{eq.low}
	\int_R^\infty \til{\rho}(r)r^2\,dr=\int_{R}^\infty v(r)^2\,dr
	\le
	R^2\til{\rho}(R)\int_R^\infty  e^{-2\mu(r -R)}\, dr =\frac{R^2}{2 \mu}\til{\rho}(R).
	\end{equation}
	Together with the upper bound, we have \eqref{eqlem.discpbound}.

	Next, we treat the case of $E=E_{N-1}<E_{N-2}$ with $N-1>Z$.
	Then \eqref{lemeq.threSchroine} becomes
	\[
	v''(r)\ge \left(\frac{N-1-Z}{r}-\frac{d}{r^2} \right)v(r).
	\]
	For $\mu < \beta < \sqrt{N-1-Z}$, the same argument as above shows
	\begin{equation}
		\label{eq.threv}
	v(r)\le v(R)e^{-2\beta (\sqrt{r}-\sqrt{R})},
	\end{equation}
	which implies
	\[
	\int_{R}^\infty v(r)^2\,dr\le v(R)^2 \left(\frac{\sqrt{R}}{2\beta} +\frac{1}{8\beta^2}\right)\le \frac{\sqrt{R}}{2\mu} v(R)^2.
	\]
	This is the desired lower bound in~\eqref{eqlem.threpbound}. The upper bound follows from the monotonicity of $v$ as above.
		\end{proof}
			The following is an immediate consequence of \eqref{eq.vbound} and \eqref{eq.threv}.
			\begin{corollary}
				For all $0<\mu<\sqrt{E_{N-1}-E}$, there exists $R_0$ such that
				\begin{equation}
				\label{eqcor.rho}
				r^2\til\rho(r)\le R^2\til{\rho}(R)e^{-2\mu(r-R)}
				\end{equation}
				for $r\ge R \ge R_0$.

				At the threshold $E=E_{N-1}<E_{N-2}$ with $N-1>Z$, we have the corresponding bound
								\begin{equation}
					\label{eqcor.threrho}
					r^2\til\rho(r)\le R^2\til{\rho}(R)e^{-4\beta (\sqrt{r}-\sqrt{R})}
				\end{equation}
				for any $0<\beta <\sqrt{N-1-Z}$ and $r\ge R\ge R_0(\beta)$.
				Moreover, $\til{\rho}(r)>0$ for $r\ge R_0$ in both cases.
				\end{corollary}

				\begin{proof}
					The pointwise bounds follow from~\eqref{eq.vbound} and \eqref{eq.threv}.
					If there is $r_0\ge R_0$ such that $\til{\rho}(r_0)=0$, by~\eqref{eqcor.rho} or~\eqref{eqcor.threrho}, $\psi =0$ on  $\{\underline{X}\colon r_N>r_0\}$.
					This contradicts the unique continuation theorem~\cite{HOStrem}.
					\end{proof}

	Let us introduce the following notation, with $z=(r\omega,\sigma)$:
	\begin{align}
		P(r)&:=r^2 \til{\rho}(r)=N\sum_{{\sigma=\uparrow,\downarrow}}\int_{\s^2}d\omega\int d\hat z_N\,\left|r\psi(\hat z_N,z)\right|^2,\\
		D(r)&:=N\sum_{{\sigma=\uparrow,\downarrow}}\int_{\s^2}d\omega\int d\hat z_N\,
		\bigl|\partial_r\bigl(r\psi(\hat z_N,z)\bigr)\bigr|^2, \\
		  S(r)&:=N\sum_{{\sigma=\uparrow,\downarrow}}\int_{\s^2}d\omega\int d\hat z_N\,|\nabla_{\s^2} \psi(\hat z_N, z)|^2,\\
		  I(r)&:=\sum_{j=2}^N \sum_{i<j}\int d\underline{z} \,\frac{\left(\mathbf{1}_{[r,\infty)}(r_i)\dfrac{x_i}{r_i}-\mathbf{1}_{[r,\infty)}(r_j)\dfrac{x_j}{r_j}\right)\cdot (x_i-x_j)}{|x_i-x_j|^3} |\psi(\underline{z})|^2,
		\end{align}
		where $\nabla_{\s^2}$ stands for the gradient with respect to $\s^2$.
		We note that $I(r)\ge 0$ since $(r_i+r_j)(1-\cos \theta)\ge0$ and $r_i-r_j\cos \theta \ge 0$ for $r_i\ge r_j$, with $x_i\cdot x_j=r_ir_j\cos \theta$.

		Our next task is to prove the following lemma.
		\begin{lemma}[Virial-type identity]
			\label{lem.virial}
			For all $R>0$, we have
			\[
			\begin{split}
			P''(R)&=4D(R)+4\int_R^\infty dr\,\frac{S(r)}{r}
			-2Z\int_R^\infty dr\,\frac{P(r)}{r^2}+2I(R).
			\end{split}
			\]
			\end{lemma}

			\begin{proof}
		Let $R_0>1$, and let $\zeta  \in C^\infty(\R)$ be a function satisfying $0\le \zeta  \le 1$ and
	\[
	\zeta (t)=1 \, \text{for } |t|>R_0,\qquad \zeta (t)=0 \, \text{ for } |t|<1.
	\]
	Following~\cite{FHHH,FH82} we introduce an antisymmetric operator
	\[
	A:=\left(\zeta(r) \frac{x}{r}\right)\cdot \nabla +\nabla \cdot \left(\zeta (r)\frac{x}{r}\right)
	=2\zeta\partial_r +\zeta'  +\frac{2}{r}\zeta
	\]
	where $\nabla$ stands for the three-dimensional gradient.
	A direct calculation shows that for real-valued $u \in H^2(\R^3)$
		\begin{align*}
		\left(u, [\Delta,A]u\right)&=2(\Delta u, A u)\\
		&=-4(\nabla u, \nabla(\zeta\partial_r u))
		-2\left(\nabla u,\nabla\left[\left(\zeta' +2\frac{\zeta}{r}\right)u\right]\right)\\
		&=-4(\partial_r u, \zeta'\partial_r u)
		-4(\nabla u, \zeta\nabla(\partial_r u))\\
		&\quad-2\left(\nabla u,\left[\zeta'+2\frac{\zeta}{r}\right]\nabla u\right)
		-2\left(u,\nabla\left[\zeta'+2\frac{\zeta}{r}\right]\cdot\nabla u\right).
		\end{align*}
		We note that
		\[
		\begin{split}
		2\left(u,\nabla\left[\zeta'+2\frac{\zeta}{r}\right]\cdot\nabla u\right)
		=\left(\nabla\left[\zeta'+2\frac{\zeta}{r}\right],\nabla u^2\right)
		=-\left(u,\left[\zeta'''+4\frac{\zeta''}{r}\right]u\right),
		\end{split}
		\]
		and, since $2\nabla u\cdot(\nabla \partial_r u)=\partial_r|\nabla u|^2+\frac{2}{r}(|\nabla u|^2-|\partial_r u|^2)$,
		\begin{align*}
		2(\nabla u,\zeta\nabla(\partial_r u))
		&=-\left(\left[\zeta'+\frac{2\zeta}{r}\right]\nabla u,\nabla u\right)
		+2\left(\frac{\zeta}{r}\nabla u,\nabla u\right)-2\left(\frac{\zeta}{r}\partial_r u,\partial_r u\right)\\
		&=-(\zeta'\nabla u,\nabla u)
		-2\left(\frac{\zeta}{r}\partial_r u,\partial_r u\right).
		\end{align*}
		Then we have
		\begin{equation}
			\label{eq.Pcomm}
		\begin{split}
			\left(u, [\Delta,A]u\right)
			&=-4(\partial_r u,\zeta'\partial_r u)
			-4\left(\frac{\zeta}{r^3}\nabla_{\s^2}u,\nabla_{\s^2}u\right)+\left(u,\left[\zeta'''+4\frac{\zeta''}{r}\right]u\right).
			\end{split}
		\end{equation}
		Next, we treat the potential $V=-Z|x|^{-1}+\sum_{i=1}^{N-1} |x_i-x|^{-1}$.
		Using $[V,A]=-2\zeta(r)\partial_r V$, where
		\[
		\partial_r V=\frac{Z}{r^2}
		-\sum_{i=1}^{N-1}\frac{x\cdot(x-x_i)}{r|x_i-x|^3},
		\]
		we obtain
		\begin{equation}
			\label{eq.Vcomm}
			\begin{split}
			(u,[V, A]u)
			&=-2Z\int_{\R^3}dx\,\frac{\zeta(r)}{r^2}u(x)^2+
			2\sum_{i=1}^{N-1}\int_{\R^3}dx\,
			\frac{x\cdot(x-x_i)}{r|x_i-x|^3}\zeta(r)u(x)^2.
			\end{split}
			\end{equation}
		From the Schr\"odinger equation $H_{N,Z}\psi = E\psi$, we see $
		0=\< \psi, [H_{N,Z},B] \psi\>
		$
	for $B$ defined by
		\[
	B:=\sum_{i=1}^N\left[\left(\zeta(r_i)\frac{x_i}{r_i}\right)\cdot\nabla_i
	+\nabla_i\cdot\left(\zeta(r_i)\frac{x_i}{r_i}\right)\right].
	\]
	Equations \eqref{eq.Pcomm} and \eqref{eq.Vcomm} lead to
	\begin{equation}
		\label{eq.Hcomm}
		\begin{split}
			\left\< \psi, [H_{N,Z}, B] \psi\right\>
			&=
			4\sum_{i=1}^N\<\partial_{r_i}\psi,\zeta'(r_i)\partial_{r_i}\psi\>
			+4\int_0^\infty dr\,\frac{\zeta(r)}{r}S(r)\\
			&\quad -\int_0^\infty dr\, \left(\zeta'''+4\frac{\zeta''}{r}\right) P(r)-2Z\int_{0}^\infty dr\,\frac{\zeta(r)}{r^2}P(r)
			\\			&\quad+2\sum_{j=2}^N \sum_{i<j}\int d\underline{z} \,\frac{\left(\zeta(r_i)\dfrac{x_i}{r_i}-\zeta(r_j)\dfrac{x_j}{r_j}\right)\cdot (x_i-x_j)}{|x_i-x_j|^3} |\psi(\underline{z})|^2.
			\end{split}
		\end{equation}
		Now fix $R>1$ and choose $R_0>R$. We take $\zeta_n \in C^\infty(\R)$ satisfying the above cutoff conditions so that $0\le \zeta_n \le 1$ and, as $n\to \infty$, $\zeta_n\to \mathbf{1}_{[R,\infty)}$ and $\zeta_n' \to \delta_R$ in the sense of distributions on $(0,\infty)$. The case $0<R\le1$ is treated by the same calculation with rescaled cutoffs.
		A limiting argument shows that, for all $R>0$,
		\begin{align*}
			&\sum_{i=1}^N\<\partial_{r_i}\psi,\zeta_n'(r_i)\partial_{r_i}\psi\>
			\to R^2
			\underbrace{N\sum_{{\sigma=\uparrow,\downarrow}}\int_{\s^2}d\omega\int d\hat z_N\,|\partial_{r_N} \psi(\hat z_N, (R\omega,\sigma))|^2}_{=:\til D(R)},\\
			&\int_0^\infty dr\,\frac{\zeta_n(r)}{r}S(r)
			\to\int_R^\infty dr\,\frac{S(r)}{r},\\
			&\sum_{j=2}^N \sum_{i<j}\int d\underline{z} \,\frac{\left(\zeta_n(r_i)\dfrac{x_i}{r_i}-\zeta_n(r_j)\dfrac{x_j}{r_j}\right)\cdot (x_i-x_j)}{|x_i-x_j|^3} |\psi(\underline{z})|^2
			\to I(R).
			\end{align*}
		Moreover, integration by parts yields
		\begin{align*}
			\int_0^\infty dr\, \left(\zeta_n'''+4\frac{\zeta_n''}{r}\right) P(r)
			&=\int_0^\infty dr\, \zeta_n'\left(P''(r)-4\frac{P'(r)}{r}+4\frac{P(r)}{r^2}\right)\\
			&\to P''(R)-4\frac{P'(R)}{R}+4\frac{P(R)}{R^2}.
			\end{align*}
			Combining these limits with \eqref{eq.Hcomm} and the identity
			\[
			R^2\til{D}(R)=D(R)-\frac{P'(R)}{R}+\frac{P(R)}{R^2},
			\]
			we have the desired conclusion.
			\end{proof}

			The above virial-type identity yields the log-convexity of $p(r)P(r)$ for a suitable positive function $p$.
			More precisely, we prove the following.

			\begin{lemma}[log-convexity inequality]
				\label{lem.logconv}
				We consider the two cases.
				\begin{enumerate}
					\item Suppose $E<E_{N-1}$. For any $0<\mu <\sqrt{E_{N-1}-E}$, it follows that
				\begin{equation}
					\label{lemeq.log}
				\frac{d^2\ln P}{dr^2}(r)\ge -\frac{Z}{\mu r^2}
				\end{equation}
				for sufficiently large $r$.
				In particular, the function $$F(r):=\ln P(r) -Z\,\frac{\ln r}{\mu}$$ is convex on $[R_0,\infty)$.
				\item Suppose $E=E_{N-1}<E_{N-2}$ and $N-1-Z>0$. For $0<\mu <\sqrt{N-1-Z}$, it follows that
				\begin{equation}
					\label{lemeq.threlog}
					\frac{d^2\ln P}{dr^2}(r)\ge -\frac{Z}{\mu r^{3/2}}
				\end{equation}
				for sufficiently large $r$.
				Then $F(r):=\ln P(r)-4Z\sqrt{r}/\mu$ is convex on $[R_0,\infty)$.
				\end{enumerate}
				\end{lemma}

				\begin{proof}
					(1) Applying the Cauchy--Schwarz inequality leads to
					\begin{align*}
					|P'(r)|^2
					&=4\left|N
					\sum_{{\sigma=\uparrow,\downarrow}}\int_{\s^2}d\omega\int d\hat z_N\,
					r\psi(\hat z_N,z) \partial_r \bigl(r\psi(\hat z_N,z)\bigr)
					\right|^2\\
					&\le 4P(r)D(r).
					\end{align*}
					Then Lemma~\ref{lem.virial} implies that
					\[
					\frac{d^2}{dr^2}\ln P(r) = \frac{P''(r)}{P(r)}-\frac{P'(r)^2}{P(r)^2}
					\ge
					-\frac{2Z}{P(r)} \int_r^\infty ds\,\frac{P(s)}{s^2}.
					\]
					From~\eqref{eqlem.discpbound}, we have
					\[
					 \int_r^\infty ds\,\frac{P(s)}{s^2}
					 \le
					 \frac{1}{r^2}
					 \int_r^\infty ds\, P(s)
					 \le\frac{P(r)}{2\mu r^2}.
					\]
					This shows \eqref{lemeq.log} and
					\begin{align*}
						F''(r)=
					\left(\ln P(r) -\frac{Z\ln r}{\mu}\right)''
					=
					\frac{P''}{P}-\frac{(P')^2}{P^2} +\frac{Z}{\mu r^2}
					\ge0,
					\end{align*}
					which proves the desired convexity.

					(2) The same reasoning shows that
					\[
					\frac{d^2}{dr^2}\ln P(r) \ge -\frac{2Z}{P(r)} \int_r^\infty ds\,\frac{P(s)}{s^2}\ge -\frac{Z}{\mu r^{3/2}}.
					\]
					This completes the proof.
					\end{proof}

					\begin{proof}[Proof of Theorem~\ref{thm.main}]
						Suppose $E<E_{N-1}$.
						We deduce from~\eqref{eqcor.rho} that for $r\ge R \ge R_0$
						\begin{equation}
							\label{eq.Fbound}
						F(r)
						\le \ln P(R) -2\mu(r-R) -\frac{Z\ln r}{\mu}.
						\end{equation}
						By the convexity of $F$, we have
						\[
						F(r)-F(R)\ge F'(R)(r-R)
						\]
						for any $r\ge R > R_0$.
						Together with~\eqref{eq.Fbound}, this also yields $F'(r)\le -2\mu$ for sufficiently large $r$.
						Since $F'$ is nondecreasing and bounded above, there exists a finite $l\le -2\mu$ such that
						\[
						\lim_{r\to \infty} F'(r)=l,\qquad \lim_{r\to \infty}\frac{F(r)}{r}=\lim_{r\to \infty}\frac{\ln P(r)}{r}=l.
						\]
						This means that
						\[
						\int_0^\infty e^{2\alpha r}P(r)\,dr
							<\infty \text{ for } 2\alpha <-l;\quad
							\int_0^\infty e^{2\alpha r}P(r)\,dr=\infty \text{ for } 2\alpha >-l.
						\]
						By definition, $\alpha_0=-l/2$ and thus
						\[
						\lim_{r\to \infty}\frac{\ln \til{\rho}(r)}{r}=-2\alpha_0.
						\]
						The above argument also works for $E=E_{N-1}<E_{N-2}$ with $N-1>Z$ by the convexity of $F(r)=\ln P(r)-4Z\sqrt{r}/\mu$ and $F'\le 0$.
						Since $(\ln \til{\rho}(r))'=F'(r)+o(1)\to -2\alpha_0$, the proof is complete.
						\end{proof}

						\section{Three-electron systems with large nuclear charges}
						\label{sect.largeZ}
In this section, we prove Theorem~\ref{thm.asymp}.
To do this, we need the following proposition, which gives a sufficient condition for $\alpha_{\underline{X}}=\alpha_0$.
\begin{proposition}
	\label{lem.equiv}
	Suppose $H_{N,Z}\psi =E\psi$ and
\begin{equation}
\label{eqthm.bind}
E<E_{N-1},\quad E+\alpha_{\underline{X}}^2 <E_{N-2}.
\end{equation}
Then $\alpha_{\underline{X}}=\alpha_0$.
	\end{proposition}

	We rely on the following theorem from~\cite[Thm.~2.1]{FH}.
	For a spherical shell $\Omega(R_1,R_2):=\{\underline{X}\in\R^{3N}\colon R_1<|\underline{X}|<R_2\}$, we use the notation
	\[
	\|\psi\|_{\Omega(R_1,R_2)}:=\left(\int_{\Omega(R_1,R_2)} |\psi(\underline{z})|^2\,d\underline{z}\right)^{1/2}.
	\]
	\begin{theorem}
		\label{thm.FH}
		For a positive function $\delta(R)$ satisfying
		\[
		\liminf_{R\to\infty} \left(\frac{\ln \delta(R)}{R}\right)\ge 0,
		\]
		it holds that $$\lim_{R\to \infty} R^{-1} \ln\|\psi\|_{\Omega(R, R+\delta(R))}^2=-2\alpha_{\underline{X}}.$$
	\end{theorem}

	\begin{proof}[Proof of Proposition~\ref{lem.equiv}.]
		Let $|x|_{N-1}:=\min_{1\le i\le N}\max_{j\neq i}|x_j|$ denote the second largest of the radii $r_1,\dots,r_N$.
		We also denote by $\Omega_R$ the region defined by
		\[
		\Omega_R:=\{\underline{X}\colon |\underline{X}|\ge(1+\delta)R, \quad |x|_\infty \le R+1\},
		\]
		where $0<\delta<1$ is a fixed constant.
		Clearly, for sufficiently large $R$, on $\Omega_R$ we have $|x|_\infty \ge (1+\delta)R/\sqrt{N}$ and $|x|_{N-1}\ge C(N,\delta)R$ since
		\[
		(1+\delta)^2R^2\le |\underline{X}|^2\le (N-1)|x|_{N-1}^2+|x|_\infty^2 \le (N-1)|x|_{N-1}^2+(R+1)^2.
		\]
		We claim the following.
		\begin{lemma}[Localization bound]
			\label{lem.localbound}
		 If $\phi \in \bigwedge^N L^2 \cap H^1$  and $\mathop{\rm supp} \phi \subset \Omega_R$, then, for sufficiently large $R$, we have
		\begin{equation}
			\label{eq.energy}
			\<\phi,H_{N,Z}\phi\> \ge \left(E_{N-2}-\frac{C}{R}\right)\<\phi,\phi\>,
			\end{equation}
			where $C=C(N,Z,\delta)$ is some constant independent of $R$.
			\end{lemma}
			\begin{proof}[Proof of Lemma~\ref{lem.localbound}]
			To prove this, we construct a partition of unity following~\cite{BFLS}.
			Let $\til f:\R_+\to \R_+$ be the nonnegative function defined by
			\[
			\til{f}(s):=\begin{cases}
				1& (0\le s\le \frac{1}{2})\\
				\cos\left(\pi\left[s-\frac{1}{2}\right]\right) & (\frac{1}{2}<s<1)\\
				0 & (s\ge 1),
				\end{cases}
			\]
			and let $f:\R_+\to\R_+$ satisfy $f(s)^2+\til{f}(s)^2=1$ for all $s\ge 0$.
			Then we define the quadratic partition of unity in a neighborhood of $\Omega_R$ by
			\[
			\Lambda_{ij}(\underline{X}):=\frac{f\left(\dfrac{|x_i|}{|x|_\infty}\right)}{\left[\displaystyle\sum_{m=1}^N f\left(\frac{|x_m|}{|x|_\infty}\right)^2\right]^{1/2}}\cdot
			\dfrac{f\left(\dfrac{|x_j|}{|\hat x_i|_\infty}\right)}{\left[\displaystyle\sum_{m\neq i}f\left(\frac{|x_m|}{|\hat{x}_i|_\infty}\right)^2\right]^{1/2}},\quad i\neq j,
			\]
			where $|\hat x_i|_\infty:=\max_{m\neq i}|x_m|$.
			Here we note that
			\[\sum_{m=1}^N f\left(\frac{|x_m|}{|x|_\infty}\right)^2\ge f(1)^2=1,\qquad
			\sum_{m\neq i} f\left(\frac{|x_m|}{|\hat x_i|_\infty}\right)^2\ge f(1)^2=1.
			\]
			By definition,
			\[
			\sum_{i=1}^N \sum_{j\neq i}\Lambda_{ij}(\underline{X})^2=1,
			\]
			and
			\begin{align*}
			\Lambda_{ij}(\underline{X})=0 \quad \text{unless }
				 |x_j|&\ge \frac{|\hat x_i|_\infty}{2},\\
				 |x_i|&\ge \frac{|x|_\infty}{2}.
			\end{align*}
			Hence, on $\Omega_R\cap\{\Lambda_{ij}\neq0\}$, we have $|x_i|, |x_j|\ge C(N,\delta)R$.
			A straightforward calculation shows that, almost everywhere,
			\begin{align*}
				\nabla_i f\left(\frac{|x_m|}{|x|_\infty}\right)
				=\begin{cases}
					|x|_\infty^{-1}f'\left(\dfrac{|x_m|}{|x|_\infty}\right)\dfrac{x_i}{|x_i|}& (i=m, |x_m|<|x|_\infty)\\[1em]
					-\dfrac{|x_m|}{|x|^2_\infty}f'\left(\dfrac{|x_m|}{|x|_\infty}\right)\dfrac{x_i}{|x_i|} & (i\neq m, |x_i|=|x|_\infty)\\[1em]
					0  & \text{otherwise}.
				\end{cases}
			\end{align*}
			Thus, since $|f'|\le\pi$ a.e., we have
			$$
			\left|\nabla_j f\left(\frac{|x_m|}{|x|_\infty}\right)\right|\le \pi |x|_\infty^{-1},\quad \left|\nabla_j f\left(\frac{|x_m|}{|\hat x_i|_\infty}\right)\right|\le \pi |\hat x_i|_\infty^{-1}$$ almost everywhere for any $j$, with $m\neq i$ in the second inequality.
			Using these estimates and the lower bounds on the denominators, we obtain, a.e. on $\Omega_R$,
			\[
			\sum_{k=1}^N\sum_{i=1}^N \sum_{j\neq i}\left| \nabla_k \Lambda_{ij}(\underline{X})\right|^2
			\le \frac{C(N,\delta)}{R^2}.
			\]
			For a given $\phi$, let $\phi_{ij}:=\Lambda_{ij}\phi$.
			Then the IMS formula~\cite{CFKS} yields
			\begin{equation}
				\label{eq.IMS}
				\<\phi,H_{N,Z}\phi\>=
				\sum_{i=1}^N \sum_{j\neq i}
				\left(
				\<\Lambda_{ij}\phi,H_{N,Z}\Lambda_{ij}\phi\>
				-\sum_{k=1}^N\int\,d\underline{z}
				\left| \nabla_k \Lambda_{ij}(\underline{X})\right|^2|\phi(\underline{z})|^2
				\right).
				\end{equation}
				By construction, $\Lambda_{ij}$ is symmetric under permutations of the variables $x_k$ with $k\neq i,j$, and therefore $\phi_{ij}$ is antisymmetric in the remaining $N-2$ variables $z_k$ for fixed $z_i,z_j$.
				On the support of $\phi_{ij}$, we have for any $i\neq j$
				\begin{equation}
					\label{eq.quad}
					\begin{split}
				H_{N,Z}&\ge H_{N-2,Z}+\sum_{k=i,j}\left(-\Delta_k -\frac{Z}{|x_k|}\right)\\
				&\ge E_{N-2}-\frac{2Z}{|x|_\infty}-\frac{2Z}{|\hat x_i|_\infty}\\
				&\ge E_{N-2} -\frac{C(N,\delta)Z}{R}
				\end{split}
				\end{equation}
				in the quadratic form sense.
				Hence we deduce from~\eqref{eq.IMS} and~\eqref{eq.quad} that
				\[
				\<\phi,H_{N,Z}\phi\>
				\ge
				\left(E_{N-2}-\frac{C(N,\delta)Z}{R}-\frac{C(N,\delta)}{R^2}\right)\<\phi, \phi\>,
				\]
				which shows the lemma.
			\end{proof}
			Now we also take smooth functions $g,h \colon \R \to [0,1]$ satisfying
			\begin{align*}
			g(s)&=0 \text{ for } s\le 0,\quad g(s)=1 \text{ for } s\ge 1;\\
			h(s)&=1 \text{ for } s\le 0,\quad h(s)=0 \text{ for } s\ge 1.
			\end{align*}
			Using these functions, we introduce
			\begin{align*}
			\chi_R(\underline{X})&:=g(|\underline{X}|-(1+\delta)R)\prod_{j=1}^N h(|x_j|-R),\\
			f_R(|\underline{X}|)&:=\alpha \min\{(|\underline{X}|-(1+\delta)R)_+, \delta R\},
			\end{align*}
			where $\alpha$ satisfies $\alpha_{\underline{X}}<\alpha<\sqrt{E_{N-2}-E}$ and $(a)_+:=\max\{a,0\}$.
			Since $e^{f_R}\chi_R \psi$ is antisymmetric and supported in $\Omega_R$, we can apply Lemma~\ref{lem.localbound}.
			By the identity~\cite[Eq.~(3.19)]{FHHH}
			\[
			\left\<
			e^{f_R}\chi_R \psi,(H_{N,Z}-E) e^{f_R}\chi_R \psi\right\>
			=\sum_{i=1}^N\<e^{f_R}\psi, |\nabla_i \chi_R+\chi_R \nabla_i f_R|^2e^{f_R}\psi\>,
			\]
			we see that, for any $\epsilon>0$, with a constant $C$ independent of $R$,
			\begin{equation}
			\label{eq.var1}
			\begin{split}
			\left(E_{N-2}-E-\frac{C}{R}\right)\|e^{f_R}\chi_R \psi\|_2^2
			&\le
			\sum_{i=1}^N\<e^{f_R}\psi, |\nabla_i \chi_R+\chi_R \nabla_i f_R|^2e^{f_R}\psi\>\\
			&\le (1+\epsilon)\alpha^2\|e^{f_R}\chi_R\psi\|_2^2\\
			&\quad+C\|\psi\|_{\Omega((1+\delta)R)}^2
			+Ce^{2\alpha\delta R}\int_R^\infty r^2\til\rho(r)\,dr,
			\end{split}
			\end{equation}
			where $\Omega(R):=\Omega(R,\infty)$ and we have used
			\begin{align*}
			\sum_{i=1}^N|\nabla_i f_R|^2&\le\alpha^2,\\
			\sum_{i=1}^N|\nabla_i\chi_R|^2
			&\le C\mathbf{1}_{\{(1+\delta)R<|\underline{X}|<(1+\delta)R+1\}}
			+C\sum_{i=1}^N\mathbf{1}_{\{R<|x_i|<R+1\}},
			\end{align*}
			and $f_R\le\alpha$ on $\{(1+\delta)R<|\underline{X}|<(1+\delta)R+1\}$.
			Choosing $\epsilon>0$ so that $(1+\epsilon)\alpha^2<E_{N-2}-E$, we obtain from~\eqref{eq.var1}, for sufficiently large $R$,
			\begin{equation}
			\label{eq.var2}
			\|e^{f_R}\chi_R\psi\|_2^2\le
			C\|\psi\|_{\Omega((1+\delta)R)}^2
			+Ce^{2\alpha\delta R}\int_R^\infty r^2\til\rho(r)\,dr.
			\end{equation}
			On the other hand, if $\delta R\ge1$, then $\chi_R=1$ and $f_R=\alpha\delta R$ on $\{|\underline{X}|\ge (1+2\delta)R,\ |x|_\infty\le R\}$. Hence we infer that
			\begin{align*}
			\|\psi\|_{\Omega( (1+2\delta)R)}^2
			&\le\sum_{i=1}^N\int_{|x_i|>R}|\psi|^2\,d\underline{z}
			+\int_{\substack{|x|_\infty\le R\\ |\underline{X}|\ge (1+2\delta)R}}|\psi|^2\,d\underline{z}\\
			&\le\int_R^\infty r^2\til\rho(r)\,dr
			+e^{-2\alpha\delta R}\|e^{f_R}\chi_R\psi\|_2^2.
			\end{align*}
			Combining this with~\eqref{eq.var2}, we have
			\[
			\|\psi\|_{\Omega( (1+2\delta)R)}^2\le C\int_R^\infty r^2\til\rho(r)\,dr
			+Ce^{-2\alpha\delta R}\|\psi\|_{\Omega((1+\delta)R)}^2.
			\]
			Taking $\delta(R)=1$ in Theorem~\ref{thm.FH} gives a lower bound for $\|\psi\|_{\Omega(R)}^2$.
			Together with the definition of $\alpha_{\underline{X}}$, this yields
			\[
			ce^{-2a\alpha_{\underline{X}} R-\epsilon R}\le\|\psi\|_{\Omega(aR)}^2\le Ce^{-2a\alpha_{\underline{X}} R+\epsilon R}
			\]
			for any fixed $\epsilon>0$ and $a>0$, and all sufficiently large $R$.
			Choosing $0<\epsilon<\delta(\alpha-\alpha_{\underline{X}})$, we obtain
			\[
			\|\psi\|_{\Omega((1+2\delta)R)}^2\le C\int_R^\infty r^2\til\rho(r)\,dr
			\]
			for sufficiently large $R$.
			Theorem~\ref{thm.main} and Lemma~\ref{lem.pbound} imply that
			\[
			\lim_{R\to\infty}\frac{1}{R}\ln\left(\int_R^\infty r^2\til\rho(r)\,dr\right)=-2\alpha_0.
			\]
			Using this limit and the above estimates, we have
			\[
			-2(1+2\delta)\alpha_{\underline{X}}\le -2\alpha_0\le -2\alpha_{\underline{X}},
			\]
			where we have used $\alpha_{\underline{X}}\le\alpha_0$.
			Letting $\delta\to0$ shows the desired conclusion.
				\end{proof}

				\begin{proof}[Proof of Theorem~\ref{thm.asymp}]
					For convenience, we rescale $x_i$ in the Hamiltonian $H_{N,Z}$ by $U=1/Z$: $\psi (\underline{z})\to \til{\psi}(\underline{z}):=Z^{-3N/2}\psi((Z^{-1}x_1, \sigma_1),\dots, (Z^{-1}x_N,\sigma_N))$ and
					\[
					\til H_{N,U}:=\til H_{N,0}+U\sum_{1\le i<j\le N}|x_i-x_j|^{-1}, \quad \til H_{N,0}:=\sum_{i=1}^N\left(-\Delta_i-\frac{1}{|x_i|}\right).
					\]
					Then $H_{N,Z}\psi=E\psi$ is equivalent to $\til H_{N,U}\til \psi=\til{E}\til{\psi}$, where $\til{E}=Z^{-2}E$.
					Moreover, $\til \alpha_0=\sup\{\alpha\colon e^{2\alpha r}{\til\rho_{\til\psi}} \in L^1(\R_+, r^2dr)\} =\alpha_0/Z$, where $\til\rho_{\til\psi}$ denotes the spherically averaged density of $\til \psi$.
					Standard perturbation theory~\cite{KatoBook,FG} tells us that as $U\to0$,
					\[
					\til E_N=e_N +O(U),
					\]
					where $e_N$ is the ground state energy of $\til H_{N,0}$.
					Let $P_0$ denote the orthogonal projection onto the eigenspace corresponding to $e_3$.
					From now on, we denote the rescaled $\til \psi$ by $\psi$.
					For the ground state $\psi$ with $\til H_{3,U}\psi =\til E_3\psi$, we can take the normalized $\phi := P_{0} \psi/\|P_{0} \psi\|_2$ so that
					\begin{equation}
						\label{eq.projcon}
					\|\psi - \phi\|_{H^2}=O(U).
					\end{equation}
					Indeed, for $Q_0=1-P_0$ we see $\|Q_0 \psi\|_{2}\le \epsilon^{-1}\|(\tilde H_{3,0}-e_3)Q_0\psi\|_2$, where $\epsilon Q_0\le \til H_{3,0}-e_3$ for some $\epsilon>0$.
					By the relative boundedness of the Coulomb potential, we get
					\begin{align*}
					\|Q_0 \psi\|_{H^2}&\le C\left(\|\til H_{3,0}Q_0\psi\|_2 +\|Q_0\psi\|_2\right)\\
					&\le
					C\left(\|(\til H_{3,0}-e_3)Q_0\psi\|_2 +(1+|e_3|)\|Q_0\psi\|_2\right)\\
					&\le C\|(\til H_{3,0}-e_3)\psi\|_2,
					\end{align*}
					where we have used $\til H_{3,0}P_0\psi=e_3 P_0\psi$.
					Using $\til{E}_3-e_3=O(U)$ and setting $W=\sum_{i<j}|x_i-x_j|^{-1}$, we obtain
					\begin{align*}
							\|Q_0 \psi\|_{H^2}&\le C\|(\til{E}_3 -e_3)\psi -UW\psi\|_2\\
							&\le CU(1+\|Q_0\psi\|_{H^2}),
						\end{align*}
						where we have used $\|W\psi\|_2\le C\|\psi\|_{H^2}\le C(1+\|Q_0\psi\|_{H^2})$.
						Hence for small $U$, we deduce from $\|Q_0 \psi\| _{H^2}=O(U)$ that
						\begin{align*}
						\|\psi - \phi\|_{H^2}
						&\le \|Q_0 \psi\|_{H^2}+\|P_0\psi-\phi\|_{H^2}\\
						&=\|Q_0 \psi\|_{H^2}+ \underbrace{\left|\|P_0 \psi\|_{2} -1 \right|}_{=O(U^2)}\|\phi\|_{H^2}\\
						&=O(U),
						\end{align*}
						which shows \eqref{eq.projcon}.

					From the spectrum of the hydrogen atom, we obtain
					\[
					e_1=-\frac{1}{4},\quad e_2=-\frac{1}{2},\quad e_3 =-\frac{9}{16}.
					\]
					Hence if we have
					\begin{equation}
						\label{eq.suffi}
					\til E_1-\til E_3 =e_1-e_3+O(U)=\frac{5}{16}+O(U)\ge \frac{\alpha_0}{Z},
					\end{equation}
					then the fact that $(\til E_1-\til E_3)^2< \til E_1-\til E_3$ for small $U$ and Proposition~\ref{lem.equiv} yield $\alpha_{\underline{X}}=\alpha_0$.
					We will prove the bound by an approximation argument.
					Recall that a basis of bound states of $-\Delta -|x|^{-1}$ is given by $\phi_{nlm}(x):=R_{nl}(r)Y_{lm}(\omega)$, where $Y_{lm}$ denote the spherical harmonics with $\int_{\s^2}|Y_{lm}(\omega)|^2\,d\omega=1$ and the radial functions $R_{nl}$ obey $\int_0^\infty|R_{nl}(r)|^2 r^2\,dr=1$, e.g.,
					\begin{equation}
						\label{eq.radial}
					R_{10}(r)=\frac{e^{-r/2}}{\sqrt{2}},\quad R_{20}(r)=\frac{4-r}{16}e^{-r/4},\quad R_{21}(r)=\frac{r}{16\sqrt{3}}e^{-r/4}.
					\end{equation}
					Since $\phi=\phi_{1s, \uparrow} \wedge \phi_{1s,\downarrow}\wedge \phi_{2\eta}$, where $\phi_{1s, \sigma'}(x,\sigma)=\phi_{100}(x)\delta_{\sigma',\sigma}$, we may write
					\begin{align*}
					\til\rho_\phi(r)&:=3\sum_{\underline{\sigma}}\int_{\s^2}\,d\omega\int_{\R^6}\,dx_1 \,dx_2|\phi(z_1,z_2,(r\omega,\sigma_3))|^2 \\
					&=2|R_{10}(r)|^2 +\sum_{{\sigma=\uparrow, \downarrow}}\int_{\s^2}\,d\omega|\phi_{2\eta}(r\omega,\sigma)|^2.
					\end{align*}
					Here $\phi_{2\eta}$ has the form $$\phi_{2\eta}(x,\sigma)=c_\sigma\phi_{200}(x)+\sum_{m=-1}^1c_{m\sigma}\phi_{21m}(x),$$ and thus, by the normalization,
					\[
					1=\underbrace{\sum_{{\sigma}}|c_\sigma|^2}_{=:a}+\sum_{m=-1}^1\sum_{{\sigma}}|c_{m\sigma}|^2.
					\]
					Then we have
					\[
					\sum_{{\sigma}}\int_{\s^2}|\phi_{2\eta}(x,\sigma)|^2 \,d\omega
					=a|R_{20}(r)|^2+(1-a)|R_{21}(r)|^2.
					\]
					Using the explicit forms in~\eqref{eq.radial}, we see
					\begin{equation}
						\label{eq.phirho}
						P_\phi(r):=r^2\til \rho_\phi(r)=r^2e^{-r}+r^2e^{-r/2}\left[\frac{a(r-4)^2}{256}+\frac{(1-a)r^2}{768}\right].
						\end{equation}
						As $r\to \infty$, we infer that
						\[
						\frac{P_\phi'(r)}{P_\phi(r)}=-\frac{1}{2}+\frac{4}{r} +O(r^{-2})
						\]
						uniformly for $a \in [0,1]$.
						Let $\til P(r):=r^2\til\rho_{\psi}(r)$.
						Using \eqref{eq.projcon}, for any $R\ge 1$, we deduce from the Sobolev inequality that
						\begin{align*}
						\sum_{\underline{\sigma}}\int_{\s^2}\int_{\R^{6}}\,dx_1dx_2\,d\omega|\partial_R\psi(z_1,z_2, (R\omega, \sigma_3))-\partial_R\phi(z_1,z_2, (R\omega, \sigma_3))|^2
						&\le C\|\psi-\phi\|_{H^2}^2\\
						&=O(U^2),
						\end{align*}
						which implies that
						\[
						\left| \til P(R)-P_\phi(R)\right|+\left| \til P'(R)-P'_\phi(R)\right|\le CR^2U.
						\]
						For each fixed $R$ and sufficiently small $U$, we have $P_\phi(R)\ge R^2e^{-R}$ and $\tilde P(R)\ge R^2e^{-R}/2$.
						Combining these bounds with the preceding estimate, we obtain
						\begin{equation}
							\label{eq.appP}
						\frac{\til P'(R)}{\til P(R)}=\frac{P_\phi'(R)}{ P_\phi(R)}+O(e^RU) =-\frac{1}{2}+\frac{4}{R}+O(R^{-2})+O(e^RU).
						\end{equation}
						Following the proof of Lemma~\ref{lem.pbound}, we see that
						\[
						\frac{d^2\sqrt{\til P}}{dr^2}(r)\ge \left(\til E_{2} -\til E_3 -\frac{1}{r}\right)\sqrt{\til P(r)}.
						\]
						By $\til E_N =e_N + O(U)$ and $e_2-e_3=1/16$, the differential inequality becomes
						\[
						\frac{d^2\sqrt{\til P}}{dr^2}(r)\ge \frac{1}{64} \sqrt{\til P(r)}
						\]
						for $r\ge 64$ and $U$ small enough.
						Hence the maximum principle leads to
						\begin{equation}
							\label{eq.phiPbound}
							\til P(r)\le \til P(R)e^{-(r-R)/4}
							\end{equation}
							for $r\ge R \ge 64$, and
							\[
							\int_R^\infty\til{P}(r)\,dr\le \frac{\til P(R)}{2\mu}
							\]
							where $\mu=1/8$.
							As in the derivation of~\eqref{lemeq.log}, we have
							\[
							(\ln \til P)''(R)\ge -\frac{2}{\til P(R)}\int_R^\infty \frac{\til P(r)}{r^2}\,dr\ge -\frac{8}{R^2},
							\]
							which shows the convexity of $F_U(r):=\ln \til{P}(r)-8\ln r$ on $r\ge 64$.
							This implies that for $R\ge 64$
							\[
							-2\til{\alpha}_0=-\frac{2\alpha_0}{Z}=\lim_{r\to \infty} F_U'(r)\ge F_U'(R)=\frac{\til P'(R)}{\til P(R)} -\frac{8}{R}.
							\]
							Together with \eqref{eq.appP}, we obtain
							\[
							-\frac{2\alpha_0}{Z}\ge -\frac{1}{2}-\frac{4}{R}+O(R^{-2})+O(e^RU),
							\]
							for $R\ge 64$.
							Letting $U\to 0$ and then $R\to\infty$, we establish
							\[
							\limsup_{Z\to \infty}\frac{\alpha_0}{Z}\le \frac{1}{4}<\frac{5}{16},
							\]
							which implies \eqref{eq.suffi}.
							Hence equality $\alpha_{\underline{X}}=\alpha_0$ follows from Proposition~\ref{lem.equiv}.

							Finally, we note that any eigenvalue $\lambda_2 >E_2$ of $H_{2,Z}$ satisfies $\lambda_2 \ge -5Z^2/16$ by the min-max principle.
							In addition, all eigenvalues $\lambda_1$ of the hydrogen atom obey $\lambda_1\ge -Z^2/4$.
							Then we infer that the threshold set satisfies $\mathcal{F}(H_{3,Z})\cap (-\infty, -5Z^2/16)=\{E_2(Z)\}$.
							From~\cite[Thm.~1.1]{FH82} and
							\[
							\frac{E_3+\alpha_{\underline{X}}^2}{Z^2}\le -\frac{9}{16}+\frac{1}{16}+o(1)<-\frac{5}{16},
							\]
							 we arrive at $E_3+\alpha_{\underline{X}}^2=E_2$.
							This completes the proof.
					\end{proof}

	\section*{Acknowledgments}
	This research is partially supported by JSPS KAKENHI Grant Number JP26K16996.
	All mathematical arguments and proofs in the manuscript were written by the author.
	ChatGPT was used to assist in checking calculations and arguments and in editing the English of the manuscript.
	The author takes full responsibility for the content.

%\section*{Statements and Declarations}
%{\bf Conflict of interest}:
%	The author declares no conflicts of interest associated with this manuscript.

%{\bf Availability of data and materials}:
%	Data sharing is not applicable to this article as no datasets were generated or analyzed during the current study.

\end{document}